\documentclass[10pt,twocolumn,showpacs,preprintnumbers,amsmath,amssymb,aps,prl,longbibliography,superscriptaddress]{revtex4-2}

\usepackage[utf8]{inputenc}
\usepackage{graphicx}
\usepackage{dcolumn}
\usepackage{bm}
\usepackage{hyperref}

\usepackage{cleveref}
\crefname{appendix}{App.}{Apps.}
\crefname{equation}{Eq.}{Eqs.}
\crefname{figure}{Fig.}{Figs.}
\crefname{table}{Tab.}{Tabs.}
\crefname{section}{Sec.}{Secs.}

\usepackage{xcolor}  
\definecolor{lightgray}{gray}{0.6}
\definecolor{medgray}{gray}{0.4}
\definecolor{mypink}{rgb}{0.858, 0.188, 0.478} 

\newcommand{\br}{\mathbf{r}}
\newcommand{\bk}{\mathbf{k}}

\newcommand{\bu}{\mathbf{u}}

\newif\ifptitle
\newif\ifpnumber
\newcounter{para}

\ptitletrue  
\pnumbertrue  

\begin{document}
\title{Inevitable First Order Phase Transitions in 3D Quantum Hall Systems}
\author{Kaiyuan Gu}
\affiliation{Department of Physics, Princeton University, Princeton, New Jersey 08544, USA}
\author{Kai Torrens}
\affiliation{Department of Physics, Harvard University, Cambridge, Massachusetts 02138, USA}
\author{Biao Lian}
\affiliation{Department of Physics, Princeton University, Princeton, New Jersey 08544, USA}
\date{\today}

\begin{abstract}
Recent experiments suggest that low carrier density three-dimensional (3D) metals ZrTe$_5$ and HfTe$_5$ exhibit the 3D quantum Hall (QH) effect with Hall resistivity plateaus and a metal-insulator transition in strong magnetic fields. The conventional 3D QH theory requires a fixed period charge density wave (CDW), which is however not observed experimentally. We investigate alternative non-CDW mechanisms by considering a 3D metal in strong magnetic fields with electrons coupled to a boson (e.g., phonon) field. We show that the model exhibits inevitable first order phase transitions at jumps of the number of occupied Landau level bands, which do not involve CDW. These transitions may drive the system into a phase separation state with percolation transitions. We further show this can lead to Hall resistivity quasi-plateaus similar to that observed experimentally, and can provide a natural explanation for the metal-insulator transition.
\end{abstract}

\maketitle

The celebrated integer quantum Hall (QH) effect of two-dimensional (2D) electron gas exhibits a quantized Hall conductance plateaus in units of $e^2/h$ \cite{klitzing1980} in terms of electron charge $e$ and Planck constant $h=2\pi\hbar$, with coefficient given by the topological Chern number which equals the number of occupied Landau levels (LLs) \cite{laughlin1981,thouless1982}. Its generalization to the 3D QH effect in 3D metals, which is topologically equivalent to infinite layers of 2D QH effects, is however extremely difficult due to the gapless bands of 3D electrons in magnetic field. A gap opening is needed for the 3D QH effect, which was first proposed to come from a fixed spatial period charge density wave (CDW) along the magnetic field \cite{Halperin1987,Montambaux1990,Kohmoto1992,Koshino2001}. Similarly, the 3D quantum anomalous Hall effect which needs no magnetic field relies on the underlying 3D periodic lattice \cite{haldane2004,burkov2011,wangj2016axion,lian2017weyl,yin2020}.

Recently, quasi-quantized plateaus of Hall resistivity was observed in low carrier density anisotropic semimetals ZrTe$_5$ and HfTe$_5$ \cite{Liu2016,wangj2018log,Tang2019,Wang2020,Galeski2020, Galeski2021,Tian2021, wu2023, Piva2024}, suggesting the potential realization of the 3D QH effect. A magnetic field independent CDW induced 3D QH effect has been proposed as a theoretical mechanism \cite{Tang2019, Wang2020,Qin2020,Geng2021}, in which the longitudinal resistivity approaches zero. However, the CDW mechanism is unsupported by other experiments showing quasi-quantized Hall resistivity but finite longitudinal resistivity \cite{Wang2020,Galeski2020, Galeski2021}, and transport and spectroscopy measurements found no evidence of CDW \cite{Galeski2021,Tian2021,Piva2024}. Even if the conventional Fermi surface Peierls instability CDW is present, its period is magnetic field dependent and cannot explain the 3D QH effect.

In this letter, we investigate the instabilities other than CDW in such 3D anisotropic metals in magnetic field. We show that when electrons interact with some boson (e.g., phonon) field, without assuming CDW, a first order phase transition (of uniform strain if the bosons are acoustic phonons) is inevitable at the transition of number of occupied Landau level (LL) bands at zero disorders. Increasing the disorders could weaken and eventually eliminate the first order transitions. Such first order phase transitions may drive the system into a phase separation state exhibiting percolation transitions, which make the Hall resistivity deviate from the linear magnetic field dependence and exhibit quasi Hall plateaus similar to the experimental observations in ZrTe$_5$ and HfTe$_5$. Our percolation theory further provides a natural explanation for the metal-insulator transition observed at large magnetic field \cite{Tang2019}, which deserves future studies. 

\begin{figure}
\centering
\includegraphics[width=0.48\textwidth]{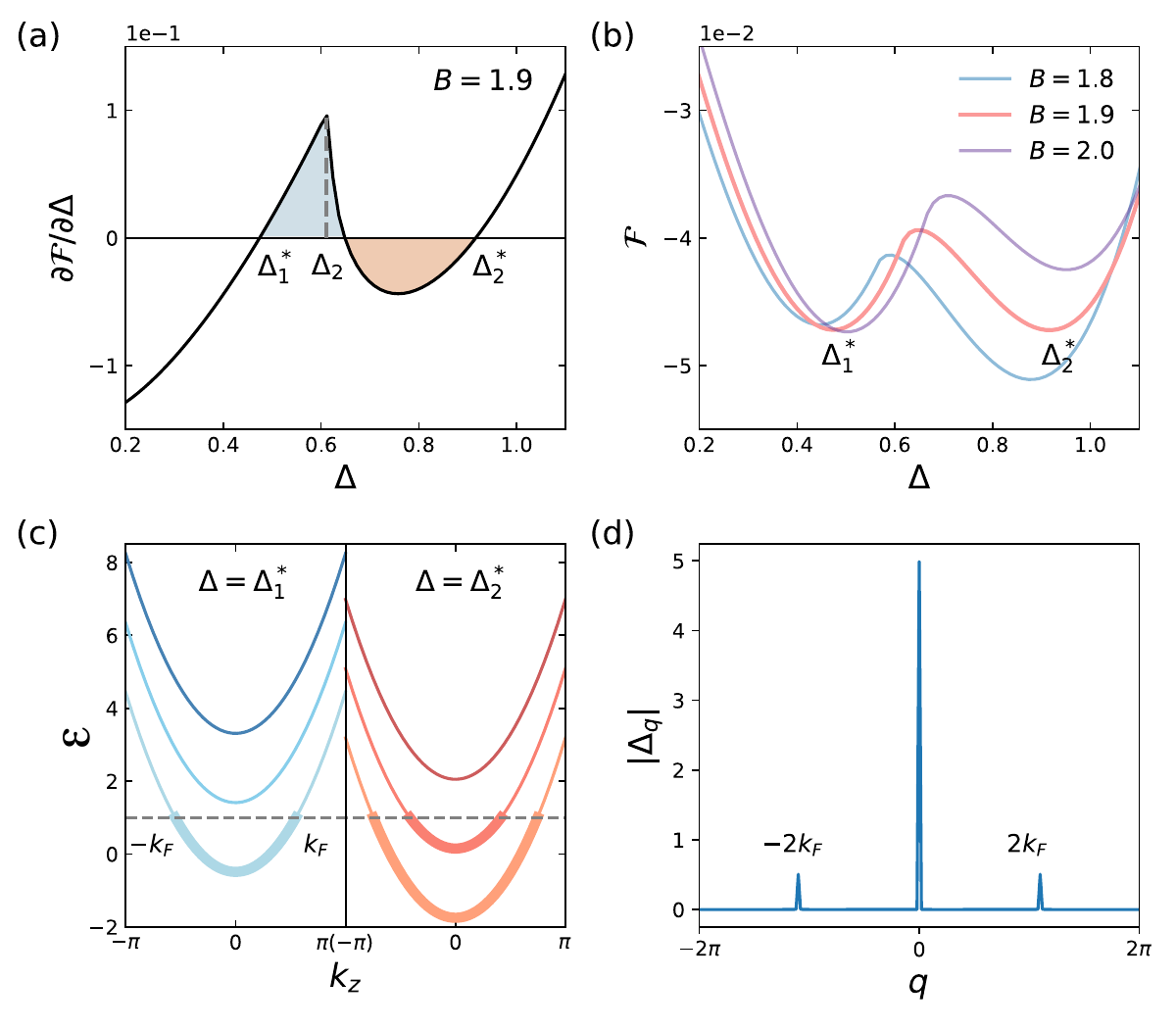}
\caption{(a) $\frac{\partial \mathcal{F}}{\partial\Delta}$ of free energy $\mathcal{F}$ at $B = 1.9$ for $D = 3$ and $\mu = 1$, which is near the transition between $N=1$ and $N=2$ phases. (b) $\mathcal{F}(\Delta)$ for different $B$, where two local minima at $\Delta_1^*$ to $\Delta_2^*$ compete, leading to a first order phase transition. (c) LL band dispersions of the two ground states at equal free energy $\mathcal{F}(\Delta_1^*) = \mathcal{F}(\Delta_2^*)$. The dashed line denotes the Fermi level. (d) Ground state calculated at $B = 2$ by keeping all $\Delta_q$, which shows a major component at $q = 0$ and small Peierls CDW components at $q = \pm 2k_F$.}
\label{fig1}
\end{figure}

\textit{The model.} Previous studies indicate that ZrTe$_5$ and HfTe$_5$ are highly anisotropic low electron density metals containing both quadratic dispersion electrons at $\Gamma$ point and massive Dirac electrons at M point \cite{Zhangy2017,Wang2020,Tian2021,wangcj2021}. For universality, we construct our model as dimensionless via a proper rescaling (see below \cref{eq:H2}), and assume a dimensionless magnetic field $B$ is applied in the $z$ direction. This yields LLs in the $x$-$y$ plane, and the $n$-th LL ($n\ge 0$, $n\in\mathbb{Z}$) has a 1D dispersion with respect to the $z$ direction dimensionless momentum $k_z$ (\cref{fig1}(c)):
\begin{equation}\label{eq:dispersion}
    \epsilon_{k_z,n}(B) = 
    \begin{cases} 
    \frac{1}{2}k_z^2 + B\Big(n+\frac{1}{2}\Big)\ ,\quad\quad\ \text{(quadratic)} \\
    M\sqrt{M^2+2Bn + k_z^2}-M^2,\text{(Dirac)}
    \end{cases}  
\end{equation}
where $M$ is the dimensionless Dirac mass gap. For simplicity, we assume only one type of electrons (quadratic or Dirac), and assume negligible spin Zeeman splitting, so that spin only contributes a factor of $2$ of degeneracy to electron numbers. Taking these simplifications or not does not qualitatively alter our theory.

Focusing on the quasi-1D interacting physics, we assume the quasi-1D LL bands in \cref{eq:dispersion} have a dimensionless Hamiltonian per volume $H=H_{\text{e}}+H_{\text{int}}+H_{\text{boson}}$, where
\begin{equation}\label{eq:H2}
    \begin{split}
        & H_{\text{e}}  =  \frac{B}{2\pi L_z}\sum_{n,k_z} \epsilon_{k_z,n}(B) \ c_{n,k_z}^\dagger c_{n,k_z}\ , \\
        & H_{\text{int}}  = - \frac{B}{2\pi L_z}\sum_{n} \int_0^{L_z} dz \ D \Delta(z) \ c_n^\dagger(z) c_n(z)\ ,\\
        & H_{\text{boson}}  = \frac{1}{L_z} \int_0^{L_z} dz \ \bigg[ \frac{1}{2}\Delta(z)^2 + \frac{\beta}{4}\Delta (z)^4 \bigg]\ , \end{split}
\end{equation}
where we have assumed a $z$-dependent mean field $\Delta(z)=\langle\hat{\Delta}(\br)\rangle$ of some dimensionless boson field $\hat{\Delta}(\br)$ coupled to the electrons. $c_{n,k_z}$ is the annihilation operator of a representative electron orbital of the $n$-th quasi-1D LL band (spin degenerate) satisfying $\{c_{m,k_{1}},c_{n,k_{2}}^\dag\}=\delta_{mn}\delta_{k_{1},k_{2}}$, and $c_n(z) = \frac{1}{\sqrt{L_z}}\sum_{k_z}e^{ik_z z}c_{n,k_z}$, with $L_z$ being the dimensionless $z$-direction system size. $H_\text{e}$ is the electron kinetic energy, $H_{\text{int}}$ is the electron-boson interaction with coupling constant $D>0$, and $H_{\text{boson}}$ is the boson field energy, in which a quartic term with $\beta>0$ is needed for the total Hamiltonian to be lower bounded.

The dimensionless model \cref{eq:H2} is from its corresponding dimensionful model rewritten in units of characteristic energy $E_0$ and lengths $a_L$, $a_z$, defined by $E_0=\zeta Y a_L^2 a_z = \hbar^2/m_L a_L^2 = \hbar^2/m_z a_z^2$ (see supplementary material (SM) \cite{suppl} Sec. I). Here $\zeta>0$ is a dimensionless number which can be chosen freely, $Y$ is the coefficient of the kinetic energy $\frac{Y}{2}\Delta'(z')^2$ of the dimensionful boson mean field $\Delta'(z')$ (primed notations denote dimensionful quantities), and $m_z$ and $m_L=\sqrt{m_xm_y}$ are the dimensionful $z$-direction and in-plane geometric mean effective Newtonian masses of electrons. For Dirac electrons with dimensionful Dirac mass $M'$ and velocities $v_{x,y,z}$, one has $m_z=M'/v_z^2$, $m_L=M'/v_xv_y$, and the dimensionless Dirac mass $M=\sqrt{M'/E_0}$. The dimensionless quantities in \cref{eq:H2} are then given by $B=eB'a_L^2/\hbar$, $D=\sqrt{2\zeta}D'/E_0$, $\beta=2\zeta\beta'/Y$, $\Delta(z)=\Delta'(z')/\sqrt{2\zeta}$, where $B'$, $D'$ and $\beta'$ are the dimensionful magnetic field, coupling constant and quartic coefficient, respectively. The free parameter $\zeta$ implies one of the dimensionless parameters in \cref{eq:H2} is redundant, but we keep $\zeta$ as a free parameter for later convenience.

The boson field may originate from acoustic/optical phonons or other emergent bosons in the system. In the case of acoustic phonons, the dimensionful boson field $\Delta'(z')=\langle \nabla'\cdot\bu\rangle$ is the bulk strain, where $\bu$ is the acoustic phonon field. Accordingly, $Y$ and $D'$ are the dimensionful bulk modulus and deformation potential, respectively.

We also define the dimensionless 3D electron density $n_{3D}=\frac{a_za_L^2}{2}n_{3D}'=\frac{B}{2\pi L_z}\sum_{n,k_z}\langle c_{n,k_z}^\dagger c_{n,k_z}\rangle$, dimensionless chemical potential $\mu=\mu'/E_0$, and dimensionless resistivity $\rho_{ij}=(e^2/ha_z)\rho_{ij}'$ ($i,j=x,y,z$), where $n_{3D}'$, $\mu'$ and $\rho_{ij}'$ are the corresponding dimensionful quantities. Accordingly, the quasi-1D bands in \cref{eq:H2} have an effective dimensionless 1D electron density $n_{1D}=\frac{1}{L_z}\sum_{n,k_z}\langle c_{n,k_z}^\dagger c_{n,k_z}\rangle=2\pi n_{3D}/B$.

Define $\Delta_q=\frac{1}{\sqrt{L_z}}\int dze^{-iqz} \Delta(z)$ as the Fourier transform of the mean field. In the Peierls instability theory, if the $n$-th quasi-1D LL band has Fermi momentum $k_{F,n}$, a nonzero CDW gap $\Delta_{\pm 2k_{F,n}}$ will develop, resulting in an insulator. Since each ``state" of the quasi-1D bands is a LL with Chern number $1$, the total Chern number of occupied states in the $x$-$y$ plane is $C_\text{tot}=\sum_{n}2k_{F,n}L_z/2\pi=n_{1D}L_z=2\pi n_{3D}L_z/B$, which leads to Hall and longitudinal resistivities
\begin{equation}\label{eq:Peierls-rho}
\rho_{xy}=\frac{L_z}{C_\text{tot}}=\frac{B}{2\pi n_{3D}}\ ,\quad \rho_{xx}=0\ .
\end{equation}
With $n_{3D}$ being a constant in the system, $\rho_{xy}\propto B$ gives no plateaus. Thus, such Peierls CDWs, if existing, are irrelevant for understanding the 3D Hall plateaus. 

\emph{First order phase transition.} As we will show below, the ground state of model in \cref{eq:H2} exhibits inevitable first order phase transitions when the number of occupied quasi-1D LL bands jumps. We minimize the zero temperature free energy $\mathcal{F}=\langle H\rangle -\mu n_{3D}$ at fixed chemical potential $\mu$, allowing the electron density $n_{3D}$ to vary. The minimization is done using the gradient descent numerical method with respect to all $\Delta_q$ (see details in SM \cite{suppl}). In the entire range of $B$ from $0$ to the quantum limit, we find a predominantly large mean field component $\Delta_0$ at $q=0$, as accompanied by small Peierls CDW components $\Delta_{q}$ at $q=\pm 2k_{F,n}$, where $k_{F,n}$ is the Fermi momentum of the $n$-th occupied quasi-1D band. An example is shown in \cref{fig1}(d). The small Peierls CDWs agree with the Peierls theory prediction $\Delta_{2k_F}\propto \exp(-2\pi^2 k_F/B)$ (see SM \cite{suppl}), which approaches zero at large $k_F/B$. 

Given the irrelevance of the small Peierls CDWs argued below \cref{eq:Peierls-rho}, and the lack of experimental evidence of CDWs \cite{Galeski2020, Galeski2021, Tian2021,Piva2024}, hereafter we ignore all the mean field components $\Delta_{q\neq 0}$, and only focus on the $q=0$ component $\Delta_0$. This amounts assuming a spatially uniform mean field $\Delta(z)=\Delta_0/\sqrt{L_z}=\Delta$. By \cref{eq:H2}, this effectively shifts the chemical potential of electrons from $\mu$ to $\mu+D\Delta$, and changes the electron density $n_{3D}$ accordingly. In the $B\rightarrow 0$ limit, where the electrons are 3D, for quadratic electron dispersion this yields a free energy
\begin{equation}\label{eq:F3D}
    \lim_{B\rightarrow0}\mathcal{F}(\Delta) = - \frac{2\sqrt{2}}{15\pi^2}(\mu + D\Delta)^{\frac{5}{2}} + \frac{1}{2}\Delta^2 + \frac{\beta}{4}\Delta^4\ .
\end{equation}
Minimizing $\mathcal{F}$ would give a spontaneous nonzero $\Delta$.

At $B>0$, we define $N$ as the number of occupied (defined as having nonzero electron density) quasi-1D LL bands, which decreases as $B$ increases. 
With interaction $D>0$, assume the $B$ field is tuned such that the free energy $\mathcal{F}$ is minimized at some $\Delta$ near $\Delta_N=\frac{\epsilon_{N,0}(B)-\mu}{D}$. In this case, the $N$-th LL band is occupied (unoccupied) when $\Delta>\Delta_N$ ($\Delta\le\Delta_N$), and has a Fermi momentum $k_{F,N}\propto\sqrt{\Delta-\Delta_N}$ if occupied. In the vicinity of $\Delta_N$, this leads to a universal form of free energy \cite{suppl}:
\begin{equation}\label{eq:Ftrans}
    \mathcal{F}(\Delta)  = \mathcal{F}_{N-1}(\Delta)-b(\Delta - \Delta_N)^{\frac{3}{2}}\Theta(\Delta-\Delta_N)\ ,
\end{equation}
where $\mathcal{F}_{N-1}(\Delta)$ is a smooth function representing the free energy of the first $N-1$ bands, $b=\sqrt{2}B D^{\frac{3}{2}}/3\pi^2$ for quadratic dispersion, and $\Theta(x)$ is the Heaviside function which is $1$ for $x>0$ and $0$ for $x\le 0$. This leads to $\frac{\partial\mathcal{F}}{\partial\Delta}\simeq c(\Delta-\Delta^*_{N-1})-\frac{3b}{2}\sqrt{\Delta - \Delta_N}\Theta(\Delta-\Delta_N)$, where $\Delta^*_{N-1}$ is the point $\mathcal{F}_{N-1}(\Delta)$ reaches its minimum, and $c=\frac{\partial^2\mathcal{F}_{N-1}}{\partial\Delta^2}>0$ is its second derivative. This implies $\frac{\partial\mathcal{F}}{\partial\Delta}$ always increases (decreases) with respect to $\Delta$ before (after) hitting $\Delta_N$, as shown in \cref{fig1}(a). Therefore, when $\Delta^*_{N-1}$ approaches $\Delta_N$ sufficiently closely from below, two competing local minima $\Delta_{N-1}^*$ and $\Delta_N^*$ with $\frac{\partial\mathcal{F}}{\partial\Delta}=0$ inevitably occur (\cref{fig1}(a),(b)), which have equal free energies $\mathcal{F}(\Delta_{N-1}^*) = \mathcal{F}(\Delta_N^*)$ when the blue and orange shaded areas in \cref{fig1}(a) are equal. Increasing $B$ thus leads to a first order phase transition of $\Delta$ jumping from $\Delta_N^*$ to $\Delta_{N-1}^*$, and the number of occupied bands jumping from $N$ to $N-1$, as shown in \cref{fig1}(b)-(c). 

\begin{figure}
  \centering
  \includegraphics[width=0.48\textwidth]{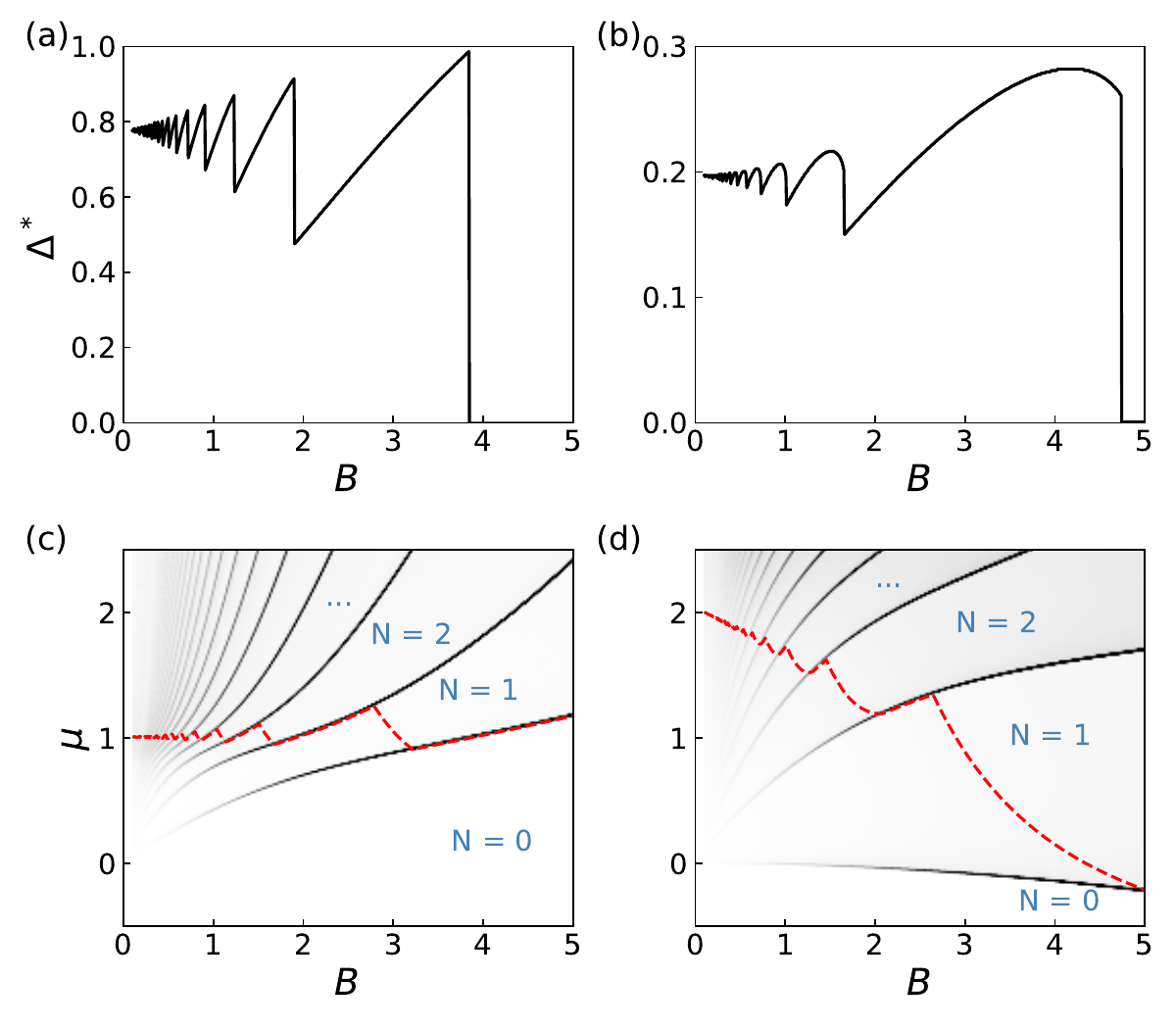}
  \caption{
  Fixing $D = 3$, $\beta = 0.2$, the ground state $\Delta^*$ with respect to $B$ calculated for (a) quadratic band with $\mu=1$, and (b) Dirac band with $M = 3$ and $\mu = 1.7 - 0.4B$. The phase diagram with the same parameters for (c) quadratic band and (d) Dirac band, with phases labeled by number of occupied LL bands $N$. The red dashed lines show the contour of $n_{3D} = 0.3$.
  }
  \label{fig2}
\end{figure}

Such first order phase transitions are inevitable as long as the LL band bottoms are quadratic at small $k_z$. To verify this, we calculate the mean field $\Delta^*(\mu,B)$ minimizing the free energy $\mathcal{F}$ of model \cref{eq:H2} with appropriate parameters estimated for ZrTe$_5$/HfTe$_5$. Hereafter, we choose the free parameter $\zeta$ to fix $a_L\simeq 25.7$nm equaling to the magnetic length of $1$ Tesla field, such that the dimensionless $B$ is simply the dimensionful magnetic field $B'$ in units of Tesla. By assuming the bulk strain from acoustic phonon field contributes a major part of the boson field $\Delta$ \cite{fub2020,kamm1985,wangcj2021,Tang2019,Jain2013}, we estimate $D=3$ as a legistimate coupling strength (see SM \cite{suppl}), and set $\beta=0.2$ (which does not sensitively affect the phase diagram). For Dirac band, $M=3$ is chosen based on \cite{Qin2020}. The experiments \cite{Tang2019,Wang2020,Galeski2020, Galeski2021} estimate a carrier density $n_{3D}'\sim 10^{17}$cm$^{-3}$, which gives $n_{3D}\sim0.3$. \cref{fig2}(a) and (b) show $\Delta^*$ as a function of $B$ at fixed chemical potential $\mu=1$ for quadratic band and $\mu=1.7-0.4B$ for Dirac band (chosen to keep $n_{3D}$ around $0.3$), respectively, which shows sharp first order jumps at the transitions of the number of occupied LL bands $N$ as expected. The jumps in $\Delta^*$ is generically larger for smaller $N$ (i.e., larger $B$). \cref{fig2}(c) and (d) show the phase diagrams for quadratic and Dirac bands, respectively, where $N$ labels the number of occupied quasi-1D LL bands.

\begin{figure}
  \centering
  \includegraphics[width=0.48\textwidth]{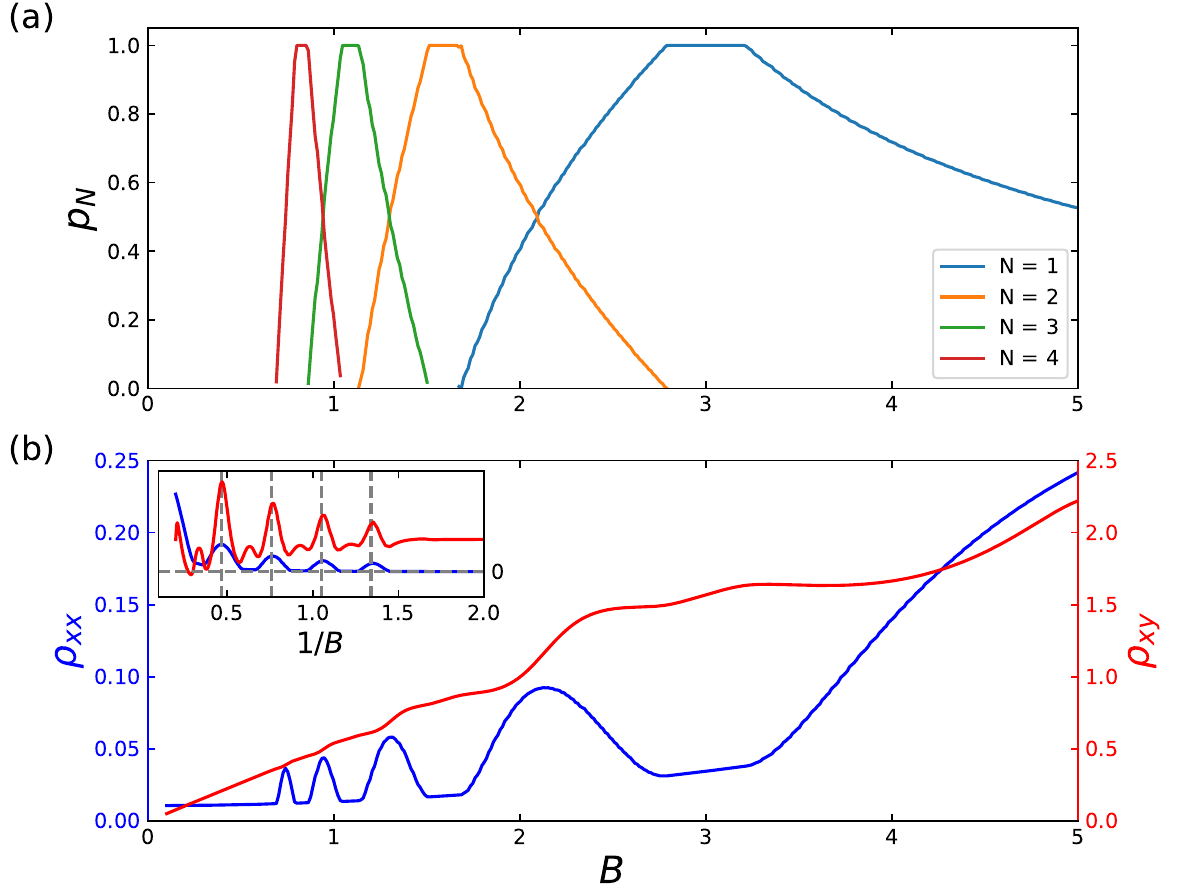}
  \caption{Along the $n_{3D}=0.3$ contour in \cref{fig2}(c), (a) spatial fractions $p_N$ of different phases $N$ up to $4$ (labels in the legend), and (b) $\rho_{xy}$ (blue) and $\rho_{xx}$ (red) with respect to $B$. The (b) inset: $\rho_{xx}$ (blue) and $\frac{Bd\rho_{xy}}{dB}$ (red) versus $1/B$.}
  \label{fig3}
\end{figure}

\textit{Phase separation and percolation.} In phase diagrams \cref{fig2}(c),(d) calculated by fixing chemical potential $\mu$, the 3D electron density $n_{3D}$ jumps across a first order phase boundary. In the physical material, $n_{3D}$ is fixed, while $\mu$ is not. Therefore, the system should evolve with $B$ along the constant $n_{3D}$ contour, as shown by the red dashed line in \cref{fig2}(c) and (d), for which $n_{3D}=0.3$. In particular, when the contour lies on the phase boundary between phases $N-1$ and $N$, the system will enter a phase separation state of both phases with spatial fractions $p_{N-1}$ and $p_{N}=1-p_{N-1}$, respectively, which is robust in 3D \cite{Imry1975}. The fractions are determined by
\begin{equation}\label{eq:n3D-ps}
n_{3D}^{(N-1)}p_{N-1}+ n_{3D}^{(N)}p_{N}=n_{3D}\ ,
\end{equation}
where $n_{3D}^{(N-1)}$ and $n_{3D}^{(N)}$ are the densities of phases $N-1$ and $N$ adjacent to the phase boundary ($n_{3D}^{(N)}>n_{3D}^{(N-1)}$). \cref{fig3}(a) shows the fractions $p_N$ calculated along the red dashed $n_{3D}=0.3$ contour in \cref{fig2}(c).

In the presence of weak disorders and random fields (varying slowly spatially),  we expect the two phases $N-1$ and $N$ to form random shaped domains in space, breaking translation symmetry in all directions. By the percolation theory in 3D, the regions of phase $N$ can percolate to infinity only if its spatial fraction $p_{N}>p_c$, and similarly for phase $N-1$, where the threshold fraction $p_c\simeq 0.3$ \cite{Isichenko1992,chalker1995,song2021}. The percolation theory is known to be crucial for understanding the Hall plateaus of 2D QH effect \cite{chalker1988,pruisken1988,huckestein1990,huo1992,wang2014}, where a partially filled LL will contribute a Hall conductance $\frac{e^2}{h}$ if percolating. For the phase $N$ in 3D here, the total number of occupied LLs in the $x$-$y$ plane is $C_\text{tot}^{(N)}=2\pi n_{3D}^{(N)}L_z/B$. Therefore, phase $N$ has $C_\text{tot}^{(N)}-C_\text{tot}^{(N-1)}$ more occupied LLs than phase $N-1$. These LLs would contribute a (dimensionless) Hall conductivity $\sigma_{xy}^{(1)}=f(p_{N})\frac{C_\text{tot}^{(N)}-C_\text{tot}^{(N-1)}}{L_z}$, where $f(p_{N})>0$ only if $p_{N}>p_c$, namely, the regions of phase $N$ percolate, thus their chiral surface states can reach the boundaries of the system, and $f(p_{N})\rightarrow1$ when $p_{N}\rightarrow 1$. The rest $C_\text{tot}^{(N-1)}$ occupied LLs possessed by both phases extend across the entire system and contribute a Hall conductivity $\sigma_{xy}^{(0)}=\frac{C_\text{tot}^{(N-1)}}{L_z}$. With \cref{eq:n3D-ps}, the total Hall conductivity $\sigma_{xy}=\sigma_{xy}^{(0)}+\sigma_{xy}^{(1)}$ is given by
\begin{equation}\label{eq:sxy}
\sigma_{xy} = \frac{2\pi}{B}\left[ n_{3D} +(n_{3D}^{(N)} - n_{3D}^{(N-1)}) \big(f(p_{N})-p_{N} \big)\right]\ .
\end{equation}

Assuming Peierls CDW gaps are absent or smaller than the disorder strength, all the phases with $N\ge1$ are metallic. This is consistent with the lack of experimental evidence of CDW \cite{Galeski2021,Tian2021,Piva2024}, and finite longitudinal resistivity $\rho_{xx}$ in experiments \cite{Wang2020,Galeski2020, Galeski2021} (except for $\rho_{xx}$ approaching zero in \cite{Tang2019}). The only exception is the $N=0$ phase, which has $n_{3D}^{(0)}=0$ and is insulating. Without knowing the microscopic details, we assume an approximate Drude model form (dimensionless) longitudinal resistivity $\rho_{xx}$:
\begin{equation}\label{eq:rxx}
\rho_{xx} = \frac{1}{4\pi n_{3D}}\bigg[\Gamma_0 (B)+\Gamma_s\big(p_N(1-p_N)\big)\bigg]
\end{equation}
where $\Gamma_0(B)=\gamma_0(1+\chi_0 B^2)$ is the dimensionless electron scattering rate typical for magneto-resistance of metals, and $\Gamma_s(p_N(1-p_N))$ is the extra scattering rate due to numerous domain walls when the system is in a phase separation state of phases $N$ and $N-1$, which is peaked at $p_N=\frac{1}{2}$. Since the jump of $\Delta^*$ across the domain walls between the two phases increases with $B$ (\cref{fig2}(a)), $\Gamma_s$ should increase with $B$, and we take an ansatz $\Gamma_s(x)=\gamma_1Bx^2$. Here $\Gamma_\alpha=\hbar \Gamma_\alpha'/E_0$ in terms of dimensionful scattering rates $\Gamma_\alpha'$ ($\alpha=0,s$). The Hall resistivity from \cref{eq:sxy,eq:rxx} is then $\rho_{xy}=\frac{1+\sqrt{1-4\rho_{xx}^2\sigma_{xy}^2}}{2\sigma_{xy}}$.

To demonstrate the result, for \cref{eq:sxy} we take $f(p)=\frac{1}{2}[1+\tanh(6p-3)]$ which is approximately nonzero only when $p>p_c\simeq0.3$, and we set $\gamma_0=0.04$, $\chi_0=0.25$ and $\gamma_1=2$, which give $\rho_{xx}$ one order smaller than $\rho_{xy}$ as is the case in experiments \cite{Tang2019,Wang2020,Galeski2020, Galeski2021}. \cref{fig3}(b) shows $\rho_{xx}$ and $\rho_{xy}$ as a function of $B$ along the constant $n_{3D}=0.3$ contour in \cref{fig2}(c). We find that $\rho_{xy}$ exhibits quasi-plateaus when the system is in a phase separation state with one of the two phases not percolating. In contrast, when the system is uniformly in one phase, $\rho_{xy}$ approaches the linear $B$ curve in \cref{eq:Peierls-rho}. The $\rho_{xy}$ plateaus are approximately equally spaced in $1/B$, as \cref{fig3}(b) inset shows. The $\rho_{xy}$ curve we obtained qualitatively resemble the experimental observations especially of HfTe$_{5}$ \cite{Galeski2020}.

Lastly, the experiments revealed a metal-insulator transition at large magnetic field \cite{Tang2019,Wang2020,Galeski2020, Galeski2021}, which can also be captured by our theory. Recall that the $N=0$ phase is insulating with $n_{3D}^{(0)}=0$. At sufficiently large $B$ ($\gtrsim 3$ in \cref{fig3}), the system enters a phase separation state with phases $N=1$ and $N=0$ coexisting, as shown in \cref{fig2}(c) and (d). As $B$ increases, the electron density $n_{3D}^{(1)}$ of phase $N=1$ increases, thus its spatial fraction $p_1$ decreases (\cref{fig3}(a)). Assume $p_1=p_c$ at magnetic field $B_c$. The $N=1$ phase is then percolating (not percolating) when $B<B_c$ ($B>B_c$), thus the entire system is metallic (insulating), leading to a metal-insulator transition at $B=B_c$. Taking $p_c=0.3$, we find $B_c\simeq8$ along the $n_{3D}=0.3$ contour in \cref{fig2}(c), the order of which matches the experiments well ($B_c'\simeq 6.7$T in \cite{Tang2019}).

\emph{Discussion}. We have shown that the 3D metal coupled with a boson field in the disorderless limit exhibits inevitable first order phase transitions in strong magnetic fields, in which the boson mean field $\Delta^*$ jumps when the number of occupied LL bands $N$ jumps. \cref{fig2}(a),(b) shows that the jump in $\Delta^*$ increases with $B$. Since the electrons feel a chemical potential shift $D\Delta^*$, if the system has short-range correlated disorders of energy scale $W$, we expect the first order phase transition to disappear when the jump in $\Delta^*$ is smaller than $W/D$. So the first order phase transition will only survive above a threshold magnetic field set by $W$.

We then showed that phase separation and percolation at the first order phase transitions give a possible non-CDW explanation of the Hall resistivity quasi-plateaus observed in 3D QH experiments in ZrTe$_5$/HfTe$_5$ \cite{Tang2019,Wang2020,Galeski2020, Galeski2021}. Moreover, the observed large magnetic field metal-insulator transition may be explained as the percolation transition of the phase separation of $N=0$ and $N=1$ phases. An interesting future questions is to investigate the critical exponents at such interacting percolation transitions. Lastly, if the boson field $\Delta^*$ originates from bulk strain, our results suggest jumps of the (dimensionful) bulk strain up to $10^{-4}$ in ZrTe$_5$/HfTe$_5$ at the first order phase transitions \cite{suppl}, and it will be intriguing if this can be measured.

\begin{acknowledgments}
\emph{Acknowledgments}. We thank Claudia Felser and Priscila Rosa for helpful discussions. This work is supported by the National Science Foundation through Princeton University’s Materials Research Science and Engineering Center DMR-2011750, and the National Science Foundation under award DMR-2141966. Additional support is provided by the Gordon and Betty Moore Foundation through Grant GBMF8685 towards the Princeton theory program.
\end{acknowledgments}

\bibliography{3DQH}




\pagebreak
\widetext
\clearpage
\begin{center}
\textbf{\large Supplemental Material for ``Inevitable First Order Phase Transitions in 3D Quantum Hall Systems"}
\end{center}
\setcounter{equation}{0}
\setcounter{figure}{0}
\setcounter{table}{0}
\setcounter{page}{1}
\makeatletter
\renewcommand{\theequation}{S\arabic{equation}}
\renewcommand{\thefigure}{S\arabic{figure}}
\renewcommand{\thesection}{\Roman{section}} 

\subsection*{I. Derivation from the Dimensionful Hamiltonian to the Dimensionless Hamiltonian}

Here we derive the dimensionless Hamiltonian in the main text Eqs. (1) and (2) from the dimensionful Hamiltonian. Assume in the absence of magnetic field, the metal has the following dimensionful 3D Hamiltonian $H'=H'_{\text{e}}+H'_{\text{int}}+H'_{\text{Boson}}$ \emph{per unit volume}:
\begin{equation}\label{eq:H0}
    \begin{split}
        & H'_{\text{e}}  =  \frac{1}{L_x'L_y L'_z}\sum_{s=\uparrow,\downarrow}\int d^3\br'   \ \psi_{s}^\dagger(\br') \mathcal{H}'(-i\nabla') \psi_{s}(\br')\ , \\
        & H'_{\text{int}}  = -\frac{1}{L_x'L_y L'_z}\sum_{s=\uparrow,\downarrow}\int d^3\br'   \ D'\hat{\Delta}'(\br')\psi_{s}^\dagger(\br') \psi_{s}(\br')\\
        & H'_{\text{boson}}  = \frac{1}{L_x'L_y L'_z} \int d^3\br' \  \bigg\{ \frac{Y}{2}\hat{\Delta}'(\br')^2 + \frac{\beta'}{4}\hat{\Delta}'(\br')^4 + \frac{1}{2}\Big[\partial_tP(\hat{\Delta}'(\br'))\Big]^2 \bigg\}
    \end{split}
\end{equation}
where all the primed notations are dimensionful physical quantities (meant to be distinguished with the corresponding dimensionless quantities defined later): $L_\alpha'$, $k_\alpha'$ ($\alpha=x,y,z$) are the system size and momentum in the $\alpha$-th direction, $\br'=(x',y',z')$ is the real space position, and $s$ is the spin. $\psi_s(\br')$ is the electron field of spin $s$, while $\hat{\Delta}'(\br')$ is the boson field coupled to it, with $D'$, $Y$ and $\beta'$ being the corresponding dimensionful parameters. $P(\hat{\Delta}'(\br'))$ is a certain function of the boson field depending on the boson theory, and the term $\frac{1}{2}\Big[\partial_tP(\hat{\Delta}'(\br'))\Big]^2$ is meant to represent the quantum dynamical term of the boson field $\hat{\Delta}'(\br')$. The kinetic energy $\epsilon'(\bk)$ of electrons, namely, the eigenvalue of the electron $\mathcal{H}'(-i\nabla')$ is assumed to take the (spin-independent) form of either quadratic or a massive Dirac (conduction band only) dispersion:
\begin{equation}\label{eq:E0}
\epsilon'(\bk')=
\begin{cases} 
\frac{\hbar^2k_x'^2}{2m_x} + \frac{\hbar^2k_y'^2}{2m_y}+\frac{\hbar^2k_z'^2}{2m_z}\ ,\qquad\qquad\quad \qquad\quad\quad\ \text{(quadratic)} \\
\sqrt{M'^2+\hbar^2(v_x^2k_x'^2+v_y^2k_y'^2+v_z^2k_z'^2)}-M',\qquad \text{(Dirac)}
\end{cases}  
\end{equation}
where $m_\alpha$ ($\alpha=x,y,z$) is the effective Newtonian mass in the $\alpha$-th direction; $M'$ is the Dirac mass (gap), and $v_\alpha$ ($\alpha=x,y,z$) is the Dirac velocity in the $\alpha$-th direction. We have set the band bottom to be at the zero energy.

We note that when the bosons are acoustic phonons, we have $\hat{\Delta}'(\br')=\nabla'\cdot \bu(\br')$ where $\bu(\br')$ is the acoustic phonon field. $Y$ and $D'$ are the bulk modulus and deformation potential of the material, respectively.

When a (dimensionful) magnetic field $B'$ is added in the $z'$-direction, Landau levels (LLs) are formed in the $x'$-$y'$ plane, and $k_z'\in 2\pi\mathbb{Z}/L_z'$ in the $z$-direction remains a good quantum number. For simplicity, we ignore the spin Zeeman splitting energy and assumes the spin is degenerate, which does not affect the main message of our paper. The spin degenerate single-electron kinetic energy of the $n$-th LL band is then given by
\begin{equation}
    \epsilon_{k_z',n}'(B') = 
    \begin{cases} 
    \frac{\hbar^2k_z'^2}{2m_z} + \frac{\hbar eB'}{m_L}\Big(n+\frac{1}{2}\Big)\ , \qquad\qquad\qquad\qquad\ \text{(quadratic)} \\
    \sqrt{M'^2+2\hbar v_xv_yeB'n+ \hbar^2 v_z^2k_z'^2}-M'\ , \qquad \text{(Dirac)}
    \end{cases}  
\end{equation}
where $m_L=\sqrt{m_xm_y}$ is the geometric mean Newtonian mass in the $x'$-$y'$ plane, and $n\ge 0$. 
As we did in the main text, we first impose translational symmetry in the $x'$ and $y'$ directions, and assume the boson field develops a time-independent mean field $\langle \hat{\Delta}'(\br') \rangle=\Delta'(z')$. We define $c^\dag_{n,k_z'}$ and $c_{n,k_z'}$ as the creation and annihilation operators of a representative electron eigenstate orbital of the $n$-th LL with $z$-direction momentum $k_z'$ and a particular spin (e.g., the eigenstate orbital at the origin of the $x'$-$y'$ plane with spin up), which satisfies $\{c_{n,k_z'},c^\dag_{n,k_z''}\}=\delta_{n,n'}\delta_{k_z',k_z''}$. Such an orbital has an area $2\pi l_B'^2=2\pi\hbar/eB'$ in the $x'$-$y'$ plane, where $l_B'=\sqrt{\hbar/eB'}$ is the magnetic length. Accordingly, for each given $n$ and $k_z$, the electron states have a LL degeneracy $N_{L}=2L_x'L_y'/2\pi l_B'^2=2L_x'L_y'eB'/2\pi\hbar$ in the $x'$-$y'$ plane, where the factor $2$ comes from spin degeneracy. The translational symmetry in the $x'$ and $y'$ directions requires each (spin degenerate) LL to be uniformly occupied/unoccupied in the $x'$-$y'$ plane, in which the number of electrons is effectively equal to $N_Lc^\dag_{n,k_z'}c_{n,k_z'}$. Thus, we can effectively rewrite the dimensionful Hamiltonian per unit volume in the mean field approximation as
\begin{equation}\label{eq:H1}
    \begin{split}
        & H'_{\text{e}}  = \frac{N_L}{L_x'L_y L'_z}\sum_{n,k'_z} \epsilon_{k_z',n}'(B') \ c_{n,k'_z}^\dagger c_{n,k'_z}=  \frac{2eB'}{2\pi \hbar L'_z}\sum_{n,k'_z} \epsilon_{k_z',n}'(B') \ c_{n,k'_z}^\dagger c_{n,k'_z}\ , \\
        & H'_{\text{int}}  = -\frac{2eB'}{2\pi \hbar L'_z}\sum _ n \int_0^{L_z'} dz' \ D' \Delta'(z') \ c_n^\dagger(z') c_n(z')\ ,\\
        & H'_{\text{boson}}  = \frac{1}{L'_z} \int_0^{L_z'} dz' \ \bigg[ \frac{Y}{2}\Delta'(z')^2 + \frac{\beta'}{4}\Delta'(z')^4 \bigg]\ ,
    \end{split}
\end{equation}
where we have defined the 1D Fourier transformed electron operator $c_n^\dag(z')=\frac{1}{\sqrt{L_z'}}\sum_{k_z'}e^{ik_z'z'} c_{n,k'_z}^\dagger$ for the $n$-th LL.

We now redefine the Hamiltonian model into a dimensionless form by rescaling. We define a characteristic energy $E_0$ and characteristic lengths $a_L$ in the $x$-$y$ plane and $a_z$ in the $z$ direction by requiring:
\begin{equation}\label{seq:scaling}
    E_0 = \zeta Y a_L^2 a_z = \frac{\hbar^2}{m_L a_L^2} = \frac{\hbar^2}{m_z a_z^2}\ ,
\end{equation}
where $\zeta>0$ is a dimensionless positive number one is free to choose (which will be chosen properly later). For quadratic electron band, $m_L=\sqrt{m_xm_y}$ and $m_z$ are simply the effective Newtonian masses. For Dirac electron band, the Newtonian masses are defined as $m_z = M'/v_z^2$ and $m_L = M'/v_xv_y$, since in the small momentum limit the dispersion in \cref{eq:E0} takes the quadratic form $\epsilon'(\bk')\simeq \frac{\hbar^2}{2M'}(v_x^2k_x'^2+v_y^2k_y'^2+v_z^2k_z'^2)$. Solving \cref{seq:scaling} yields the following expressions:
\begin{equation}\label{seq:solving-scaling}
a_L=\left(\frac{\hbar^2m_z^{1/2}}{\zeta Ym_L^{3/2}}\right)^{1/5}\ ,\qquad a_z=\left(\frac{\hbar^2m_L}{\zeta Ym_z^{2}}\right)^{1/5}\ ,\qquad E_0=\left(\frac{\hbar^6\zeta^2Y^2}{m_L^2m_z}\right)^{1/5}\ .
\end{equation}
We then define the following dimensionless quantities (which do not have prime in notation):
\begin{equation}\label{seq:rescaling}
\begin{split}
&H=\frac{H'}{2\zeta Y}\ ,\quad L_z=\frac{L_z'}{a_z}\ ,\quad z=\frac{z'}{a_z}\ ,\quad k_z=k_z'a_z\ ,\quad B=\frac{a_L^2}{l_B'^2}=\frac{ea_L^2}{\hbar}B'\ ,\quad \epsilon_{k_z,n}(B)=\frac{\epsilon_{k_z',n}'(B')}{E_0}\ ,\  M=\sqrt{\frac{M'}{E_0}}\ , \\
& c^\dag_{n,k_z}=c^\dag_{n,k_z'}\ ,\quad c_n^\dagger(z)=\frac{1}{\sqrt{L_z}}\sum_{k_z}e^{ik_zz} c_{n,k_z}^\dagger=\sqrt{a_z}c_n^\dagger(z')\ ,\quad \Delta(z)=\frac{\Delta'(z')}{\sqrt{2\zeta}}\ ,\quad \beta=\frac{2\zeta\beta'}{Y}\ ,\quad D=\frac{\sqrt{2\zeta}D'}{E_0}\ .
\end{split}
\end{equation}
In addition, assume $\mu'$ and $n_{3D}'$ are the dimensionful physical chemical potential and 3D electron density, respectively. We define the dimensionless chemical potential $\mu$ and dimensionless electron density $n_{3D}$ as
\begin{equation}\label{seq:chem-dens}
\mu=\frac{\mu'}{E_0}\ ,\qquad n_{3D}=\frac{n_{3D}'a_L^2a_z}{2}\ ,
\end{equation}
After such a rescaling, the dimensionless electron dispersions becomes
\begin{equation}\label{seq:dispersion}
    \epsilon_{k_z,n}(B) = 
    \begin{cases} 
    \frac{1}{2}k_z^2 + B\Big(n+\frac{1}{2}\Big)\ ,\quad\quad\qquad\qquad \text{(quadratic)} \\
    M\sqrt{M^2+2Bn + k_z^2}-M^2, \qquad\  \text{(Dirac)}
    \end{cases}  
\end{equation}
where for the Dirac band the dimensionless mass is $M=M'a_z/\hbar v_z=\sqrt{M'/E_0}$ as define in \cref{seq:rescaling}. The dimensionless Hamiltonian per unit volume $H=H_{\text{e}}+H_{\text{int}}+H_{\text{boson}}$ after the rescaling takes the form
\begin{equation}\label{seq:H2}
    \begin{split}
        & H_{\text{e}}  =  \frac{B}{2\pi L_z}\sum_{n,k_z} \epsilon_{k_z,n}(B) \ c_{n,k_z}^\dagger c_{n,k_z}\ , \\
        & H_{\text{int}}  = - \frac{B}{2\pi L_z}\sum_{n} \int_0^{L_z} dz \ D \Delta(z) \ c_n^\dagger(z) c_n(z)\ ,\\
        & H_{\text{boson}}  = \frac{1}{L_z} \int_0^{L_z} dz \ \bigg[ \frac{1}{2}\Delta(z)^2 + \frac{\beta}{4}\Delta (z)^4 \bigg]\ , \end{split}
\end{equation}\label{seq:dens}
as given in the main text. The dimensionless 3D electron density in \cref{seq:chem-dens} satisfies
\begin{equation}
n_{3D}=\frac{B}{2\pi L_z}\sum_{n,k_z} \ \langle c_{n,k_z}^\dagger c_{n,k_z}\rangle \ .
\end{equation}

In particular, we note that the form of the dimensionless model in \cref{seq:H2} is independent of the choice of the dimensionless number $\zeta$ in \cref{seq:scaling}. This implies that the three dimensionless parameters $B$, $D$ and $\beta$ are not independent: we can set one of the parameters to $1$ by choosing a number proper $\zeta$. However, we keep all the three parameters for convenience.

We now take the estimated physical parameters for ZrTe$_5$ and HfTe$_5$, and estimate the dimensionless parameters here, which we use for numerical calculations. For this purpose, we assume the boson field $\Delta'=\langle \nabla'\cdot \bu(\br')\rangle$ is from acoustic phonon field $\bu(\br')$, and thus $D'$ is the deformation potential, and $Y$ is the bulk modulus. In principle, the other boson fields such as optical phonons can also contribute, which we assume are of similar order of magnitudes. From the literature \cite{fub2020,kamm1985,wangcj2021,Tang2019,Qin2020,Jain2013}, we adopt the following order estimations of the physical parameters:
\begin{equation}
D'\sim 10\text{eV}\ ,\quad Y\sim 5\times10^{9}\text{Pa}\simeq 3.1\times 10^{28}\text{eV}\cdot\text{m}^{-3}\ ,\quad m_L\simeq 0.2 m_e\ ,\quad m_z\simeq 2m_e\ ,\quad M'\simeq 5\text{meV}\ ,
\end{equation}
where $m_e\simeq 0.91\times 10^{30}$kg is the bare electron mass, which yields $\hbar^2/m_e\simeq 1.22\times 10^{-38}$J$\cdot$m$^2=7.63\times 10^{-20}$eV$\cdot$m$^2$. Allowing $\zeta$ to be a free parameter, this yields
\begin{equation}
a_L\simeq 0.52\zeta^{-1/5}\text{nm}\ .
\end{equation}
For convenience, we choose $\zeta$ to set $a_L$ to be equal to the magnetic length $\sqrt{\hbar\times 1\text{T}/e}=25.7$nm for a 1 Tesla magnetic field, which gives:
\begin{equation}
a_L\simeq 0.52\zeta^{-1/5}\text{nm}\simeq 25.7\text{nm}\ ,\qquad\rightarrow\qquad  \zeta\simeq 3.4\times 10^{-9}\ .
\end{equation}
In this way, the dimensionless magnetic field $B$ is simply the dimensionful magnetic field $B'$ measured in Tesla, which we consider as a convenient choice. 
Under this choice, we derive the characteristic scales in \cref{seq:solving-scaling} and dimensionless parameters in \cref{seq:rescaling} to be
\begin{equation}
a_L\simeq 25.7\text{nm}\ ,\quad a_z\simeq 8.1\text{nm}\ ,\quad E_0\simeq 0.5\text{meV}\ ,\quad D= \frac{\sqrt{2\xi}D'}{E_0}\sim 2\ ,\quad M=\sqrt{\frac{M'}{E_0}}\sim 3\ .
\end{equation}
In the experiment, the 3D electron density is around $n_{3D}'\sim 10^{17}$cm$^{-3}=10^4$nm$^{-3}$, which gives the dimensionless density
\begin{equation}
n_{3D}=\frac{n_{3D}'a_L^2a_z}{2}\sim 0.3\ .
\end{equation}

Considering that the coupling to the other boson fields may further enhance the effective coupling constant $D$, we choose 
\begin{equation}
D=3\ ,\qquad n_{3D}=0.3
\end{equation}
for our calculations. The parameter $\beta$ of the quartic boson field term is not reported in literature, and is solely for the purpose to lower bound the Hamiltonian. For small $\beta$, our calculations are not sensitive to $\beta$. The value of $\beta=0.2$ chosen in our calculations belong to this range.

An additional note: from the above derivations, $E_0\simeq 0.5$meV gives the magnitude of the dimensionful LL spacings at physical magnetic field $B'\simeq 1$T. At the same time, the spin Zeeman splitting is on the order of $\mu_B B'\simeq 0.1$meV for $B'\simeq 1$T. This justifies our simplifying approximation that the spin Zeeman splitting is ignorable compared to the orbital LL spacing.

\subsection*{II. Numerical Calculation of ground states with respect to all $\Delta_{q}$} 

In this section, we show the numerical calculation of the ground states of the system by optimizing the free energy $\mathcal{F}(\Delta_{q})=\langle H\rangle-\mu n_{3D}$ at fixed $\mu$ with respect to all $\Delta_q$, where $\Delta_q$ are the Fourier components of $\Delta(z) = \frac{1}{\sqrt{L_z}}\sum_{q} \Delta_{q} e^{iqz}$. We rewrite the dimensionless Hamiltonian (\cref{seq:H2}) in momentum space:

\begin{equation}
    \begin{split}
        & H_{\text{e}}  =  \frac{B}{2\pi L_z}\sum_{n,k_z} \epsilon_{k_z,n}(B) \ c_{n,k_z}^\dagger c_{n,k_z}\ , \\
        & H_{\text{int}}  = -\frac{B}{2\pi L_z}\sum_{n,k_z,q} \frac{D\Delta_{q}}{\sqrt{L_z}}c_{n,k_z+q}^\dagger c_{n,k_z} \\
        &  H_{\text{boson}}  = \frac{1}{L_z}\bigg[ \frac{1}{2}\sum_{q}\Delta_{q}\Delta_{-q} + \frac{\beta}{4L_z}\sum_{q_1,q_2,q_3} \Delta_{q_1}\Delta_{q_2}\Delta_{q_3}\Delta_{-(q_1+q_2+q_3)}\bigg]
    \end{split}
\end{equation}
We allow all $\Delta_q$ to be complex in our calculations, which are required to satisfy $\Delta_q=\overline{\Delta}_{-q}$, where $\overline{c}$ stands for the complex conjugate of $c$.

Denote $\Psi_n = (\{c_{n,k_z}\})^T$ as the base of $n^{th}$ Landau band, then the Hamiltonian subtracting $\mu \hat{n}_{3D}$ ($\hat{n}_{3D}$ is the 3D electron density operator) can be written as

\begin{equation}
    H-\mu \hat{n}_{3D}= \frac{B}{2\pi L_z}\sum_{n}\Psi_n^\dagger \big(\mathcal{H}_e+\sum_q \mathcal{H}_{q}-\mu\big) \Psi_n + \sum_{q}\Delta_{q} \Omega_{q}
\end{equation}

where $\mathcal{H}_e$ is a diagonal matrix for electron energy term $\varepsilon_0$, and $\mathcal{H}_{q}$ accounts for the electron-boson mean field coupling, including the uniform term at $q = 0$, and 

\begin{equation}
    \Omega_{q} = \frac{1}{2L_z}\Delta_{-q} + \frac{\beta}{4L_z^2}\sum_{q_1+q_2+q_3 = -q}\Delta_{q_1}\Delta_{q_2}\Delta_{q_3}\ .
\end{equation}

In our numerical calculation, We sample $k_z$ with the standard spacing $2\pi/L_z$. For each configuration $\Delta_{q}$, we diagonalize $\mathcal{H} = \mathcal{H}_e+\sum_{q} \mathcal{H}_{q}$, and calculate the total energy by summing all the occupied electron states with eigenvalues smaller than $\mu$ for each Landau band $n$, after which we get the free energy $\mathcal{F}(\Delta_{q})$. Then, we iterate using the gradient descent method to minimize the the free energy for each iteration: in the $j$-th iteration step, we update the boson mean field by gradient descent method, $\Delta_{q}^{(j)} = \Delta_{q}^{(j-1)} - \gamma \frac{\delta \mathcal{F} (\Delta_{q}^{(j-1)})}{\delta \overline{\Delta}_q}$ with a suitable length of step $\gamma$, which would converge to the ground state or some local minimum. By choosing sufficiently many initial configurations for iteration, we select the lowest final state as the ground state. 

\begin{figure}
  \centering
  \includegraphics[width=0.9\textwidth]{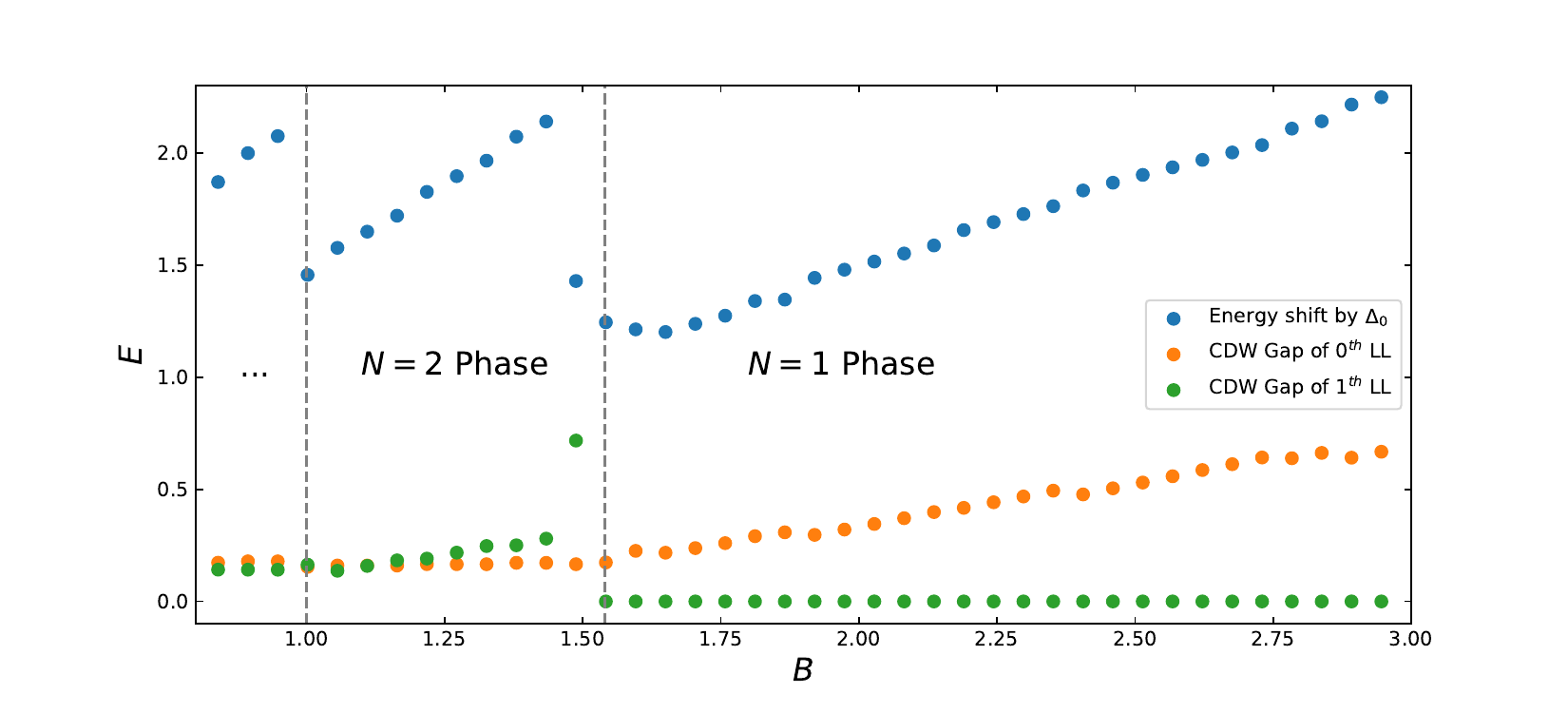}
  \caption{Comparison of energy shift by $\Delta_0$ and the CDW gap for different $B$.}
  \label{figs1}
\end{figure}

For quadratic Landau bands with $L_z = 100$, $\mu = 1$, $B = 2$, $D = 3$ and $\beta = 0.2$, the result is shown in the main text Fig. 1(d), where both $\Delta_0$ and the Peierls CDW gap $\Delta_{2k_F}$ are found to be nonzero, but $|\Delta_0|$ is much larger than $|\Delta_{2k_{F,n}}|$. For different $B$, the amplitudes of the two peaks are shown in \cref{figs1}. We note that the CDW gaps $\Delta_{2k_{F,n}}$ are generically much smaller than $\Delta_0$ except when $k_{F,n}$ is very small, namely, close to the first order phase transitions. However, when $k_{F,n}$ is very small, $\Delta_{2k_{F,n}}$ has almost zero momentum $q=2k_{F,n}\approx 0$, which is not quite distinguishable from $\Delta_0$, so we consider them playing the same role as giving an almost uniform mean field $\Delta(z)=\Delta$. This justifies our approximation of keeping only $\Delta_0$ in the main text.

We can also further verify the above results and the smallness of the Peierls CDW gaps by analytical calculations in the quantum limit. Assume there is only one LL band occupied, and only $\Delta_0$ and $\Delta_{2k_F}$ are non-zero. We also neglect the $\beta$ quartic term of the boson energy. This goes back to the conventional Peierls theory. Then, for a fixed $\Delta_0$, the free energy as a function of $\Delta_{2k_F}$ is

\begin{equation}
    \mathcal{F}(\Delta_{2k_F}) \approx -\frac{B}{2\pi}\int_{-k_F}^{k_F} \frac{dk}{2\pi} \sqrt{\Big(\varepsilon_0(k)-\mu-\frac{D\Delta_0}{\sqrt{L_z}}\Big)^2 + \bigg(\frac{D\Delta_{2k_F}}{\sqrt{L_z}}\bigg)^2} + \frac{1}{L_z}|\Delta_{2k_F}|^2\ .
\end{equation}
Minimizing the free energy by $\partial  \mathcal{F}(\Delta_{2k_F})/ \partial \Delta_{2k_F} = 0$ gives
\begin{equation}
    \Delta_{2k_F} \approx \frac{k_F^2}{2\sinh \big( \frac{2\pi^2k_F}{B}\big)}
\end{equation}
which decreases exponentially when $B$ is small, $\Delta_{2k_F} \propto \exp(-2\pi^2k_F/B) \rightarrow 0$ for sufficiently large $k_F$. When $B$ is large, this CDW gap can be eventually comparable to $\Delta_0$.

\subsection*{III. {Free energy with respect to uniform $\Delta$}} 
Here we derive the free energy $\mathcal{F}(\Delta)$ of the dimensionless Hamiltonian (at fixed chemical potential $\mu$) by assuming uniform boson mean field $\Delta(z)=\Delta$, based on which we calculate the phase diagrams in the main text by minimizing $\mathcal{F}(\Delta)$.

For constant $\Delta(z) = \Delta = \Delta_0/\sqrt{L_z}$, the dimensionless free energy of the system is

\begin{equation}
    \mathcal{F}(\Delta,B)  =  \frac{B}{2\pi L_z}\sum_{n\in\text{occ},k_z} \big[\epsilon_{k_z,n}(B) -  D \Delta - \mu \big] +  \frac{1}{2}\Delta^2 + \frac{\beta}{4}\Delta^4
\end{equation}
where the summation in $n$ and $k_z$ is within the range $\epsilon_{k_z,n}(B)\le D \Delta + \mu$ (i.e., occupied bands). Define $k_{F,n}\ge0$ as the Fermi momentum of the $n$-th LL band satisfying $\epsilon_{k_{F,n},n}(B)= D \Delta + \mu$. When $L_z \rightarrow \infty$, we have
\begin{equation}
    \mathcal{F}(\Delta,B) =  \frac{B}{2\pi}\sum_{n\in\text{occ}} \int_{-k_{F,n}}^{k_{F,n}} \frac{dk_z}{2\pi} \ \big[\epsilon_{k_z,n}(B) -  D \Delta - \mu \big] +  \frac{1}{2}\Delta^2 + \frac{\beta}{4}\Delta^4\ ,
\end{equation}
where $n\in\text{occ}$ runs over all the occupied bands (with $\epsilon_{0,n}(B)\le D \Delta + \mu$).

(i) For the quadratic band where $\epsilon_{k_z,n}(B) = \frac{1}{2}k_z^2 + B(n+\frac{1}{2})$, by substituting into variable $E=\frac{1}{2}k_z^2$ and integration by part, 
\begin{equation}
\begin{split}
    \mathcal{F}(\Delta,B) & =  - \frac{B}{2\pi}\sum_{n\in\text{occ}} \frac{2}{2\pi}\int_0^{\mu + D\Delta-B(n+1/2)} \sqrt{2E} dE +  \frac{1}{2}\Delta^2 + \frac{\beta}{4}\Delta^4 \\
    & = - \frac{\sqrt{2}}{3\pi^2}B \sum_{n\in\text{occ}}\bigg[\mu + D\Delta - B\big(n+\frac{1}{2}\big)\bigg]^{3/2} + \frac{1}{2}\Delta^2 + \frac{\beta}{4}\Delta^4\ ,
\end{split}
\end{equation}
and the corresponding density $n_{3D}$ is
\begin{equation}
    n_{3D}(\Delta,B) = -\frac{\partial \mathcal{F}(\Delta,B)}{\partial \mu} = \frac{\sqrt{2}}{2\pi^2}B\sum_{n\in\text{occ}}\bigg[\mu + D\Delta - B\big(n+\frac{1}{2}\big)\bigg]^{1/2}\ .
\end{equation}
When $B \rightarrow 0$, 
\begin{equation}
    \lim_{B\rightarrow 0 } \mathcal{F}(\Delta,B) = - \frac{2\sqrt{2}}{15\pi^2}(\mu + D\Delta)^{5/2} + \frac{1}{2}\Delta^2 + \frac{\beta}{4}\Delta^4\ .
\end{equation}

and

\begin{equation}
    \lim_{B\rightarrow 0 } n_{3D}(\Delta,B) = \frac{\sqrt{2}}{3\pi^2}(\mu + D\Delta)^{3/2}
\end{equation}

(ii) For the Dirac band where $\epsilon_{k_z,n}(B) = M\sqrt{M^2+2Bn + k_z^2}-M^2$, similarly we have

\begin{equation}
\begin{split}
    \mathcal{F}(\Delta,B) & =  - \frac{B}{2\pi}\sum_{n\in\text{occ}} \frac{2}{2\pi}\int_{M(\sqrt{M^2+2Bn}-M)-D\Delta}^{\mu} \sqrt{\big(\frac{E}{M}+M \big) - M^2 - 2Bn} \ dE +  \frac{1}{2}\Delta^2 + \frac{\beta}{4}\Delta^4 \\
    & = - \frac{B}{2\pi^2}\sum_{n\in\text{occ}} \int_{\sqrt{M^2+2Bn}}^{\frac{\mu+D\Delta}{M}+M} \sqrt{E'^2 - M^2 - 2Bn} \ dE' + \frac{1}{2}\Delta^2 + \frac{\beta}{4}\Delta^4 \\
    & = - \frac{B}{4\pi^2}\sum_{n\in\text{occ}}(M^2+2Bn) \bigg[x\sqrt{x^2-1} - \ln(x+\sqrt{x^2-1})\bigg] + \frac{1}{2}\Delta^2 + \frac{\beta}{4}\Delta^4\ , \\
\end{split}
\end{equation}
where
\begin{equation}
    x = \frac{\frac{\mu+D\Delta}{M}+M}{\sqrt{M^2 + 2Bn}}\ ,
\end{equation}
and the 3D density is
\begin{equation}
    n_{3D}(\Delta,B) = \frac{B}{2\pi^2}\sum_{n\in\text{occ}} \sqrt{M^2+2Bn}\sqrt{x^2-1}\ .
\end{equation}

\end{document}